 \documentclass[smallabstract,smallcaptions]{dccpaper}

\usepackage{epsfig}
\usepackage{citesort}
\usepackage{amsmath}
\usepackage{amssymb}
\usepackage{color}
\usepackage{url}
\usepackage{booktabs}
\usepackage{subcaption}
\usepackage{enumitem}
\usepackage{cite}
\usepackage{multirow}
\usepackage{graphicx}
\usepackage{makecell}
\usepackage{pifont}
\newlength{\figurewidth}
\newlength{\smallfigurewidth}

\begin{document}

\title
{\large
\textbf{DeepFGS: Fine-Grained Scalable Coding for Learned Image Compression}
}

\author{%
Yongqi Zhai$^{1,2\ast}$, Yi Ma$^{1,3\ast}$, Luyang Tang$^{1,2}$, Wei Jiang $^{1}$ and Ronggang Wang$^{1,2\dag}$\\[0.5em]
{\small\begin{minipage}{\linewidth}\begin{center}
\begin{tabular}{c}
$^{1}$Guangdong Provincial Key Laboratory of Ultra High Definition Immersive \\Media Technology, Peking University Shenzhen Graduate School, China \\
$^{2}$Pengcheng Laboratory, China \\
$^{3}$Huawei Technologies Co., Ltd, China \\
\url{zhaiyongqi@stu.pku.edu.cn,}
\thinspace
\url{rgwang@pkusz.edu.cn}
\end{tabular}
\end{center}\end{minipage}}
}

\maketitle
\thispagestyle{empty}

\begin{abstract}
Scalable coding, which can adapt to channel bandwidth variation, performs well in today's complex network environment. However, most existing scalable compression methods face two challenges: reduced compression performance and insufficient scalability. To overcome the above problems, this paper proposes a learned fine-grained scalable image compression framework, namely DeepFGS. Specifically, we introduce a feature separation backbone to divide the image information into basic and scalable features, then redistribute the features channel by channel through an information rearrangement strategy. In this way, we can generate a continuously scalable bitstream via one-pass encoding. For entropy coding, we design a mutual entropy model to fully explore the correlation between the basic and scalable features. In addition, we reuse the decoder to reduce the parameters and computational complexity. Experiments demonstrate that our proposed DeepFGS outperforms previous learning-based scalable image compression models and traditional scalable image codecs in both PSNR and MS-SSIM metrics.
\end{abstract}

\renewcommand{\thefootnote}{}
\footnote{$\ast$ Both authors contributed equally. $\dag$ Ronggang Wang is the corresponding author.}

\Section{Introduction}
Image coding technology aims to remove redundant information in images and compress image data dozens of times. With the coming of the metaverse, massive amounts of image data are generated and transmitted. Therefore, image coding technologies have become more and more important. Recently, learning-based image compression has developed rapidly, and the most advanced model \cite{jiang2023mlic} has completely surpassed the traditional codecs in terms of PSNR and MS-SSIM.

Nevertheless, when we turn our focus from the pursuit of performance to actual application scenarios, learning-based image compression still lacks the realization of an important function: scalable bitstream.
The bitstream is scalable if a subset of the bitstream can also generate a useful representation, i.e. the decoder can selectively decode part of the bitstream depending on the bandwidth. There are some works trying to realize learned scalable coding. For example, \cite{jia2019layered} uses a layered coding structure to make the bitstream coarse-grained scalable (the bitstream has four independently decodable subsets), but due to the use of multiple codecs, the inferencing process is complicated, and the scalability of the bitstream is limited. The RNN-based method \cite{toderici2015variable,toderici2017full,johnston2018improved} can generate a coarse-grained scalable bitstream through multiple iterations. However, the complexity of the encoding-decoding process is too high, and the rate-distortion performance is lower than that of CNN-based methods.

\begin{figure*}
  \centering
  \includegraphics[width=0.65\linewidth, trim = {90 40 90 20}, clip]{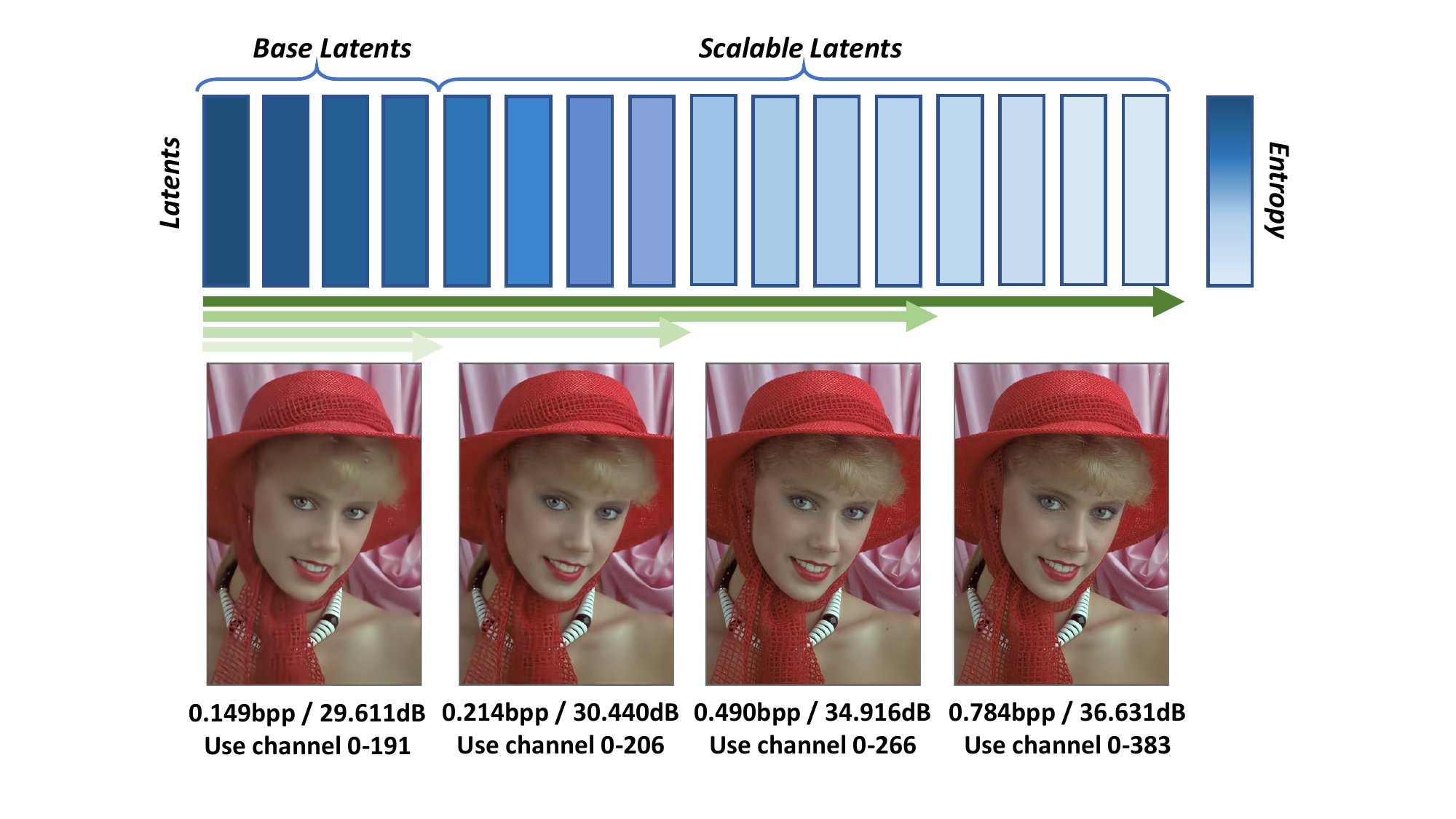} 
  \caption{Illustration of fine-grained scalable Coding of the proposed DeepFGS.}
  \label{brief_framework}
  \vspace{-0.5cm}
\end{figure*}

To achieve fine-grained scalable image coding, this paper proposes a novel framework, namely DeepFGS, which can provide a highly flexible bitstream with only one encoding process to cover an entire bitrate range. It means that even if the bitstream is truncated at any position, the truncated bitstream can be reconstructed into a complete image.
The generation of our scalable bitstream requires the following steps: First, the image is decomposed into basic features and scalable features through the feature separation backbone, and the redundancy between the two features is eliminated from pixel-level to feature-level. Then, the information rearrangement strategy redistributes the scalable features and establishes the forward dependence between channels instead of two-way dependence to adapt to the scalable decoding process. In this way, each additional decoded channel will bring continuous quality gain. 
Figure \ref{brief_framework} visualizes the process of fine-grained scalable coding of DeepFGS. As we can see, a single bitstream can be truncated anywhere and decoded at different quality levels.
In addition, to reduce the error in the entropy estimation process, we improve the entropy model with the mutual information of the basic and scalable features.
Experimental results show that our model has excellent scalability, and each subset of the bitstream can decode images of different quality. Moreover, the rate-distortion performance of our model is better than previous scalable compression methods. 

Our contributions can be summarized as follows:

\begin{itemize}[topsep=1pt, itemsep=1pt, parsep=1pt]
\item[\small\textbullet] We propose a fine-grained scalable coding framework for learned image compression, namely DeepFGS, which  outperforms previous scalable methods in both PSNR and MS-SSIM metrics.
\item[\small\textbullet] We introduce a feature separation backbone and information rearrangement strategy to obtain an extremely flexible bitstream, which can provide a continuously adjustable image quality.
\item[\small\textbullet] We design a mutual entropy model, which uses mutual information between the basic and scalable features to obtain more accurate probability estimation.
\item[\small\textbullet] To reduce the parameters and computational complexity, we design a feature fusion module and reuse the decoder.
\end{itemize}

\begin{figure*}[t]
  \centering
  \includegraphics[width=0.93\linewidth, trim = {10 10 10 10}, clip]{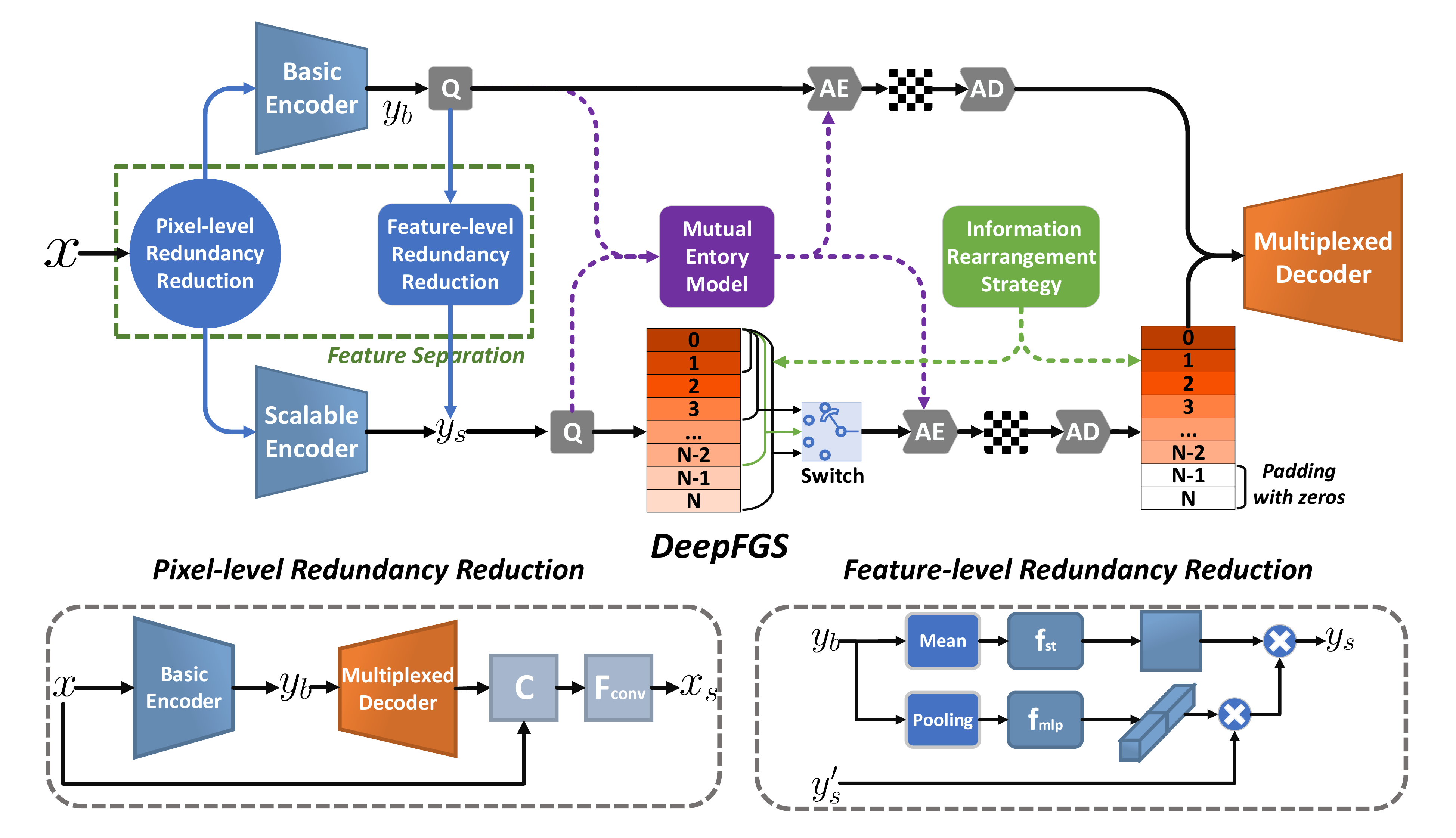}
  \caption{The architecture of our fine-grained scalable coding framework DeepFGS. Q denotes quantization. AE and AD are arithmetic encoder and decoder.}
  \label{framework}
  \vspace{-0.5cm}
\end{figure*}

\Section{Proposed Method}
As shown in Figure \ref{framework}, the feature separation backbone divides the image into basic features $y_b$ and scalable features $y_s$. Then through the information rearrangement strategy, the information in the latent representation is redistributed. We have deployed a feature selection switch before the entropy coding and a feature fusion module in the front of the decoder to achieve scalable coding and decoder reuse. 

\SubSection{Feature Separation Backbone}

To achieve scalable coding, only a certain subset of the bitstream can be used to reconstruct the image, which will cause the image quality to be spatially uneven. To solve this issue, we divide the features of the image $x$ into two parts. The first is the basic features $y_b$ that contribute to the indispensable basic texture of the image, and the second part is the scalable features $y_s$ that bring optional quality improvement. We reduce the redundancy between $y_b$ and $y_s$ from pixel domain to feature domain. Specifically, as shown in Figure \ref{framework}, image $x$ is sent to the basic encoder $g_b$ to extract the basic features $y_b$. Then, we input $y_b$ to the decoder $g_d$ to get the basic reconstructed image $\hat{x}_b$. The scalable features $y_{s'}$ will be obtained by: 
\begin{align}
&   y_{s'} =g_s(F_{conv}(x || \hat{x}_b)),
\label{equation2}
\end{align}
where $||$ denotes the concat operation, $F_{conv}$ denotes the convolution operation that is used to calculate the residual between $x$ and $\hat{x}_b$, and $g_s$ denotes the scalable encoder. 

In order to guide the basic layer and the scalable layer to learn different levels of features (e.g., the basic layer learns more low-frequency information, and the scalable layer learns more texture information), we design a feature-level redundancy reduction module (FRR). Inspired by the spatial channel attention mechanism \cite{woo2018cbam}, we use $y_b$ as a cross guidance to filter the redundant information in the $y_s$ in the channel and space domain. Figure \ref{framework} shows the architecture of our feature-level redundancy reduction module, which can be formulated as follows:
\begin{align}
   y_{s} =y_{s'} \otimes f_{mlp}(pooling(y_{b})) \otimes f_{st}(mean(y_{b})).
\label{equation3}
\end{align}
The $pooling$ operation and the $mean$ operation refer to the average in channel and space respectively, serving the channel-spatial attention mechanism. $\otimes$ denotes the channel-wise multiplication, $f_{mlp}$ denotes multi-layer perceptron, and $f_{st}$ denotes the operation that first unsqueezes the channel and then squeezes it to 1.
In this way, we have obtained compact latent representations $y_b$ and $y_s$, and then we will redistribute the scalable features $y_s$ through the information rearrangement strategy.

\begin{figure*}
  \centering
  \includegraphics[width=\linewidth]{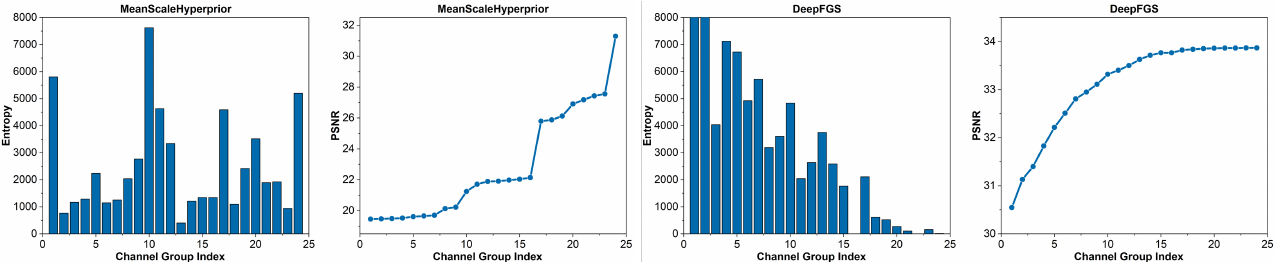} 
  \caption{Visualization of the entropy and the reconstruction quality of latent representation from kodim20. We divide the 192 channels into 24 groups on average, and the figure shows the entropy and the reconstruction quality of each group.}
  \label{entropy}
  \vspace{-0.5cm}
\end{figure*}

\SubSection{Information Rearrangement Strategy}
We first conduct two simple experiments to demonstrate that the contribution of latent representation of non-scalable models to reconstruction is not channel-wise uniform. Taking the image $x$ as input to the encoder (MeanScaleHyperprior \cite{minnen2018joint}), we can obtain its latent representation $y$. We divide the channels of $y$ (192 channels in total) equally into 24 groups, for example, 0-7, 8-15, $\cdots$, 184-191. Then, we calculated the entropy of each group of channels. As shown in Figure \ref{entropy}, the information distribution of the non-scalable image compression network is messy and irregular. We also use the first n groups (the remaining groups are padding with zeros) to reconstruct the image, the PSNR curve is not smooth and has jumps. So we can conclude that the information distribution of the non-scalable image compression model is disordered and contributes unevenly to the reconstruction quality in terms of channels.


To achieve continuous and smooth growth in quality, we use the information rearrangement strategy to redistribute the scalable features $y_s$ and establish the forward dependence between channels instead of two-way dependence, which means that each channel only needs to be combined with the previous channel to complete the reconstruction. Instead of explicitly rearranging features, we designed an optimization function to guide network learning on how to do so.
In our framework, the features to be encoded include $y_b \in \mathbb{R}^{C_1 \times H \times W}$ and $y_s \in \mathbb{R}^{C_2 \times H \times W}$, both $C_1$ and $C_2$ is set to 192. Let $k\in [C_1, C_1+C_2]\cap \mathbb{N}$ denote the number of channels available for decoding, up to 384, which simulates bandwidth fluctuations. When $k$ is equal to $C_1$, it means that only the basic features $y_b$ are available. Therefore, the optimization function is:
\begin{equation}
L_{basic} =R(y_b)+\lambda D(x,g_d(y_b)),
\label{equation4}
\end{equation}
$D$ denotes the distortion metric, $R$ refers to the bitrate, and $\lambda$ is used to balance the rate-distortion tradeoff. When $k$ is greater than 192, additional scalable features can be obtained, and we need to optimize all forward-dependent channel combinations:
\begin{align}
\label{equation_j}
 &  L_{scalable} = \sum_{i=1}^{k-C_1} R(y_b) +R(y_s^{\leq i}) + \lambda w(i) D(x,g_d(y_b || y_s^{\leq i})),
\end{align}
where $y_s^{\leq i}$ refers to the first $i$ channels in $y_s$, $w(i) = \lfloor i / 8 \rfloor$ provides variable distortion weights for different feature combinations. We simplify the training process by replacing the accumulation operation with the sampling operation:
\begin{align}
 &  L_{scalable} = R(y_b) +R(y_s^{\leq j}) + \lambda w(j) D(x,g_d(y_b || y_s^{\leq j})),
\label{equation6}
\end{align}
where $j$ is sampled from $[C_1,C_2]\cap \mathbb{N}$.
Figure \ref{entropy} illustrates the role of information rearrangement strategy, it can be observed that our model arranges the channels in descending entropy order. As to the quality of the reconstructed image, the non-scalable model is volatile when using the first $n$ groups for decoding, while our model maintains a smooth and steady growth in an extensive bit rate range. Note that the information rearrangement strategy is performed during the training phase. After model training, features are rearranged channel by channel based on the importance of information. Therefore, there is no need to transmit side information to identify the importance of information during the encoding (inference) phase.
\begin{figure*}[t]
  \centering
  \includegraphics[width=0.62\linewidth, trim = {80 280 80 220}, clip]{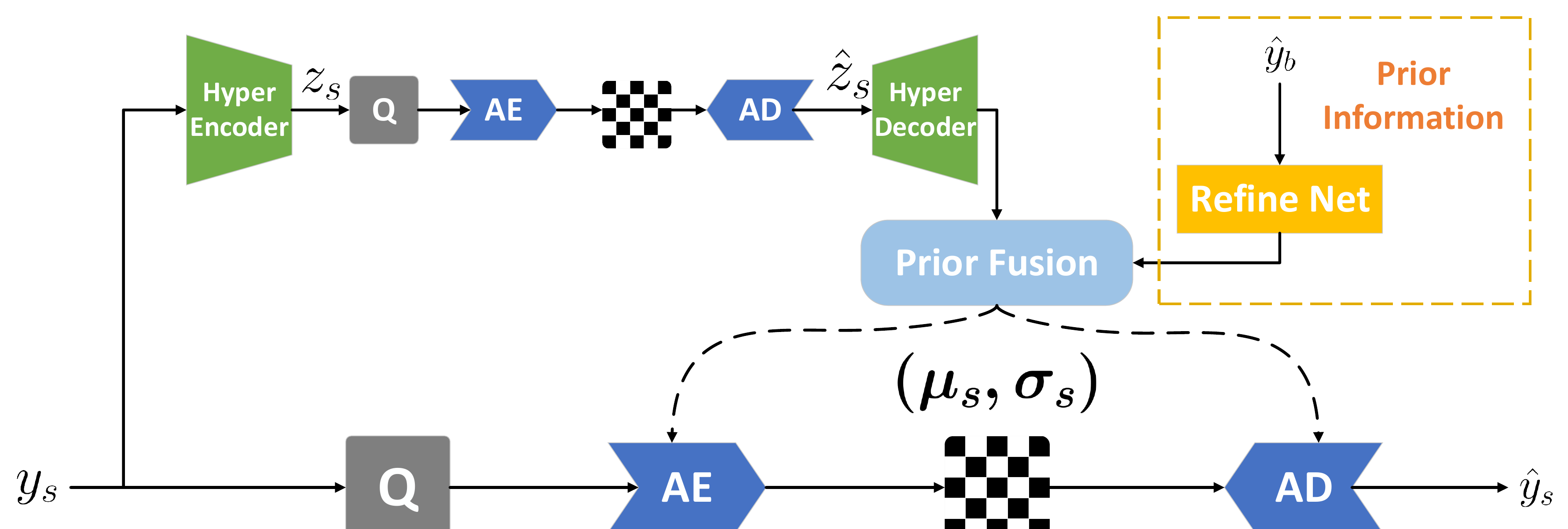}
  \caption{The architecture of our mutual entropy model for $\hat{y}_s$.}
  \label{MEM}
  \vspace{-0.3cm}
\end{figure*}

\SubSection{Mutual Entropy Model}
\label{Entropy_Model}
In the previous methods \cite{jia2019layered,mei2021learning}, they use independent entropy models for the basic and enhancement layers. However, due to the existence of redundancy between latents, the mutual information ${I(\hat{y}_s, \hat{y}_b)}$ is positive. Let ${H(\hat{y}_s)}$ denotes the Shannon entropy of $\hat{y}_s$, then the conditional entropy ${H(\hat{y}_s|\hat{y}_b)}$ can be obtained as:
\begin{align}
  H(\hat{y}_s|\hat{y}_b) &= H(\hat{y}_s) - I(\hat{y}_s, \hat{y}_b) < H(\hat{y}_s).
  \label{ih}
\end{align}
Therefore, in our entropy model, $\hat{y}_s$ is conditioned on the $\hat{y}_b$. With this prior, we can estimate the probability of $\hat{y}_s$ more accurately, thus improving the R-D performance. Figure \ref{MEM} shows the architecture of our mutual entropy model.

\begin{figure*}
  \centering
  \includegraphics[width=0.71\linewidth, trim = {0 100 0 102}, clip]{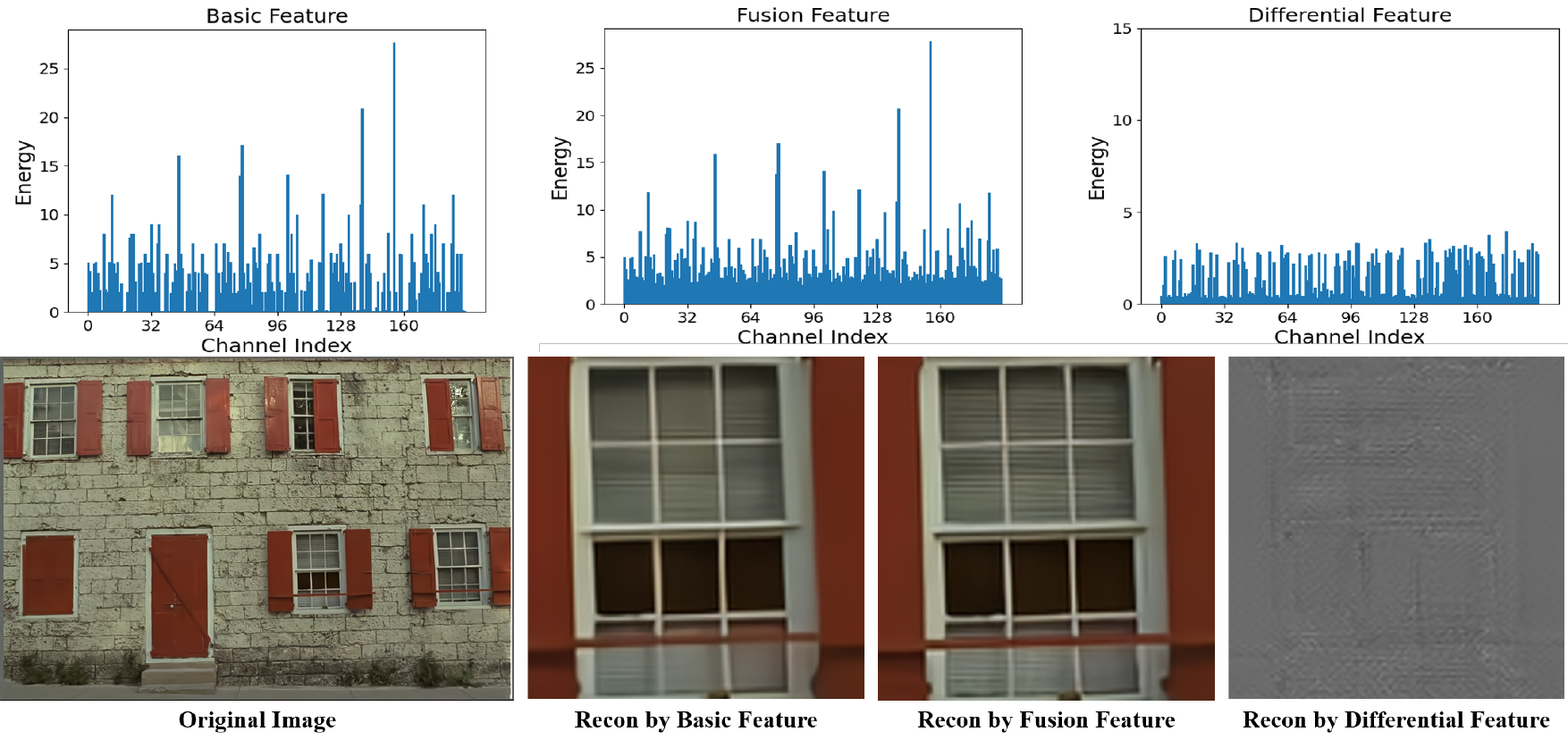} 
  \caption{Observation of the process of feature fusion. Basic features, fusion features, and difference features refer to: $FFM(\hat{y}_b)$, $FFM(\hat{y}_b || \hat{y}_s)$ and $FFM(\hat{y}_b || \hat{y}_s) - FFM(\hat{y}_b)$. We search for the feature map with the largest activation value among the features and visualize the energy of each channel ($maximum-minimum$).}
  \label{view_feature}
  \vspace{-0.3cm}
\end{figure*}

\SubSection{Decoder Multiplexing}
\label{Decoder_reuse}
Many studies on learned scalable image compression, such as \cite{jia2019layered} and \cite{mei2021learning}, generate bitstreams of different quality levels, which use more than two decoders to decode these bitstreams. Considering the deployment of multiple decoders will increase the number of parameters and training difficulty of the model, therefore, the complexity can be reduced by multiplexing part of the decoder parameters. As shown in Figure \ref{framework}, our model includes two reconstruction processes, which are located in the feature separation backbone and at the end of the entire framework. In order to reduce the parameters and complexity of the decoder network, the two locations share the same decoder. Since the
basic layer features and enhancement layer features have different properties, simple concatenates cannot take advantage of their complementarity. Therefore, we design a feature fusion module (FFM) at the head of the decoder to better fuse features $\hat{y}_b$ and $\hat{y}_s$. The FFM module has the same structure as the FRR module and the fusion features $\hat{y}_m$ is obtained:
\begin{align}
  \hat{y}_{m}=\hat{y}_{d} \otimes f_{mlp}(pooling(\hat{y}_{d})) \otimes f_{st}(mean(\hat{y}_{d})), \quad \hat{y}_{d} =Switch(\hat{y}_b, \hat{y}_b || \hat{y}_s).
\label{equation7}
\end{align}

Figure \ref{view_feature} visualizes the process of feature fusion. From the first row, we can observe that the energy distribution of the basic features and the fusion features are similar. The difference (differential features) between the two is reflected in some low-energy channels. By comparing with the reconstructed image in the second row, we can see that the high-energy channels of the basic features carry low-frequency basic textures. The differential features include high-frequency textures that can improve the quality of the reconstruction, which is consistent with our expectations for the feature separation backbone.

\Section{Experiments}
\SubSection{Training Details}
COCO \cite{lin2014microsoft} dataset is used for training, and all images are randomly cropped into ${256 \times 256}$ patches during training. We adopt Adam \cite{kingma2014adam} optimizer with a batch size of 16 to train our model. Our network is optimized for 50 epochs with an initial learning rate of $1{e^{-4}}$, and the learning rate is reduced to $1{e^{-5}}$ for the last 15 epochs. We use $\lambda$ to adjust the minimum bitrate and use the function $w(\cdot)$ to adjust the bitrate range covered by the model. During training, $k$ is randomly selected as an integer in [192,384] to change the size of the bitstream received by the decoder. Our models are optimized with two quality metrics, i.e. MSE and MS-SSIM \cite{wang2003multiscale}. For the MSE models, $\lambda$ is set to 0.002. For the MS-SSIM models, $\lambda$ is set to 7.0.

\begin{figure*}[t]
    \centering
  \begin{subfigure}[]{0.42\textwidth}
        \centering
      \includegraphics[width=1\textwidth]{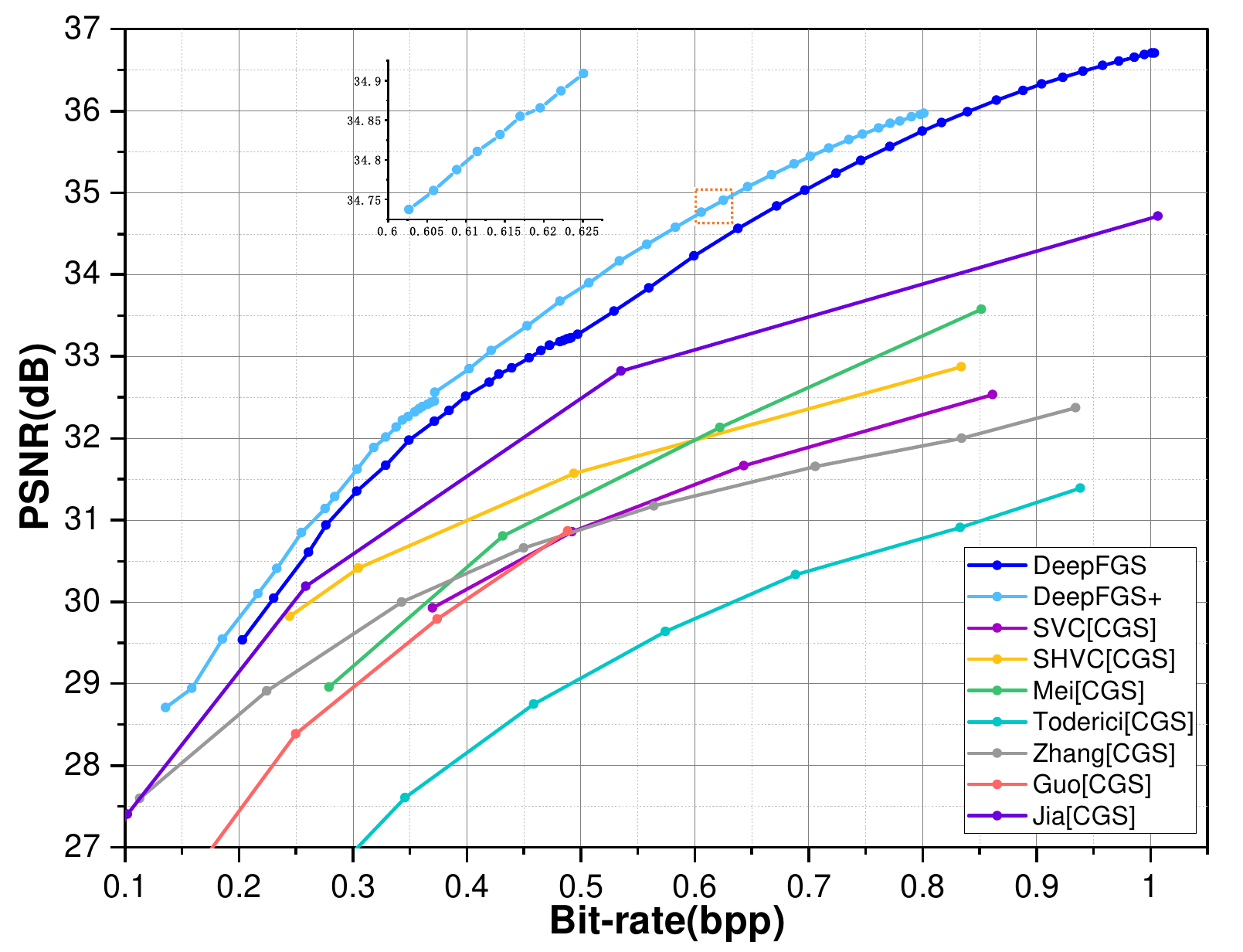} %
        \caption{}
        \label{subfigure_1}
    \end{subfigure}
    \hspace{10mm}
  \begin{subfigure}[]{0.42\textwidth}
        \centering
      \includegraphics[width=1\textwidth]{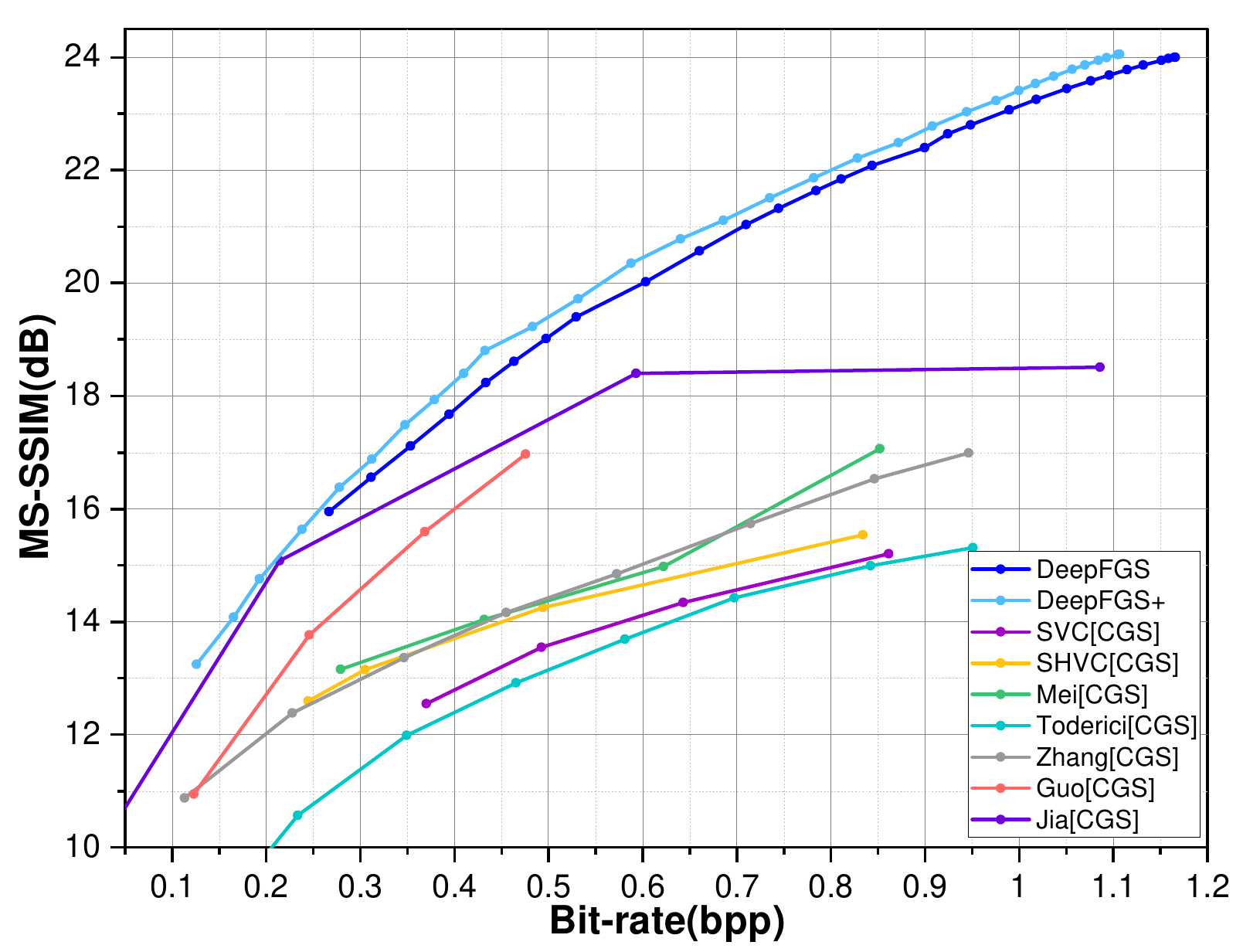}
        \caption{}
        \label{subfigure_2}
    \end{subfigure}
    \hfill
  \begin{subfigure}[]{0.42\textwidth}
        \centering
      \includegraphics[width=1\textwidth]{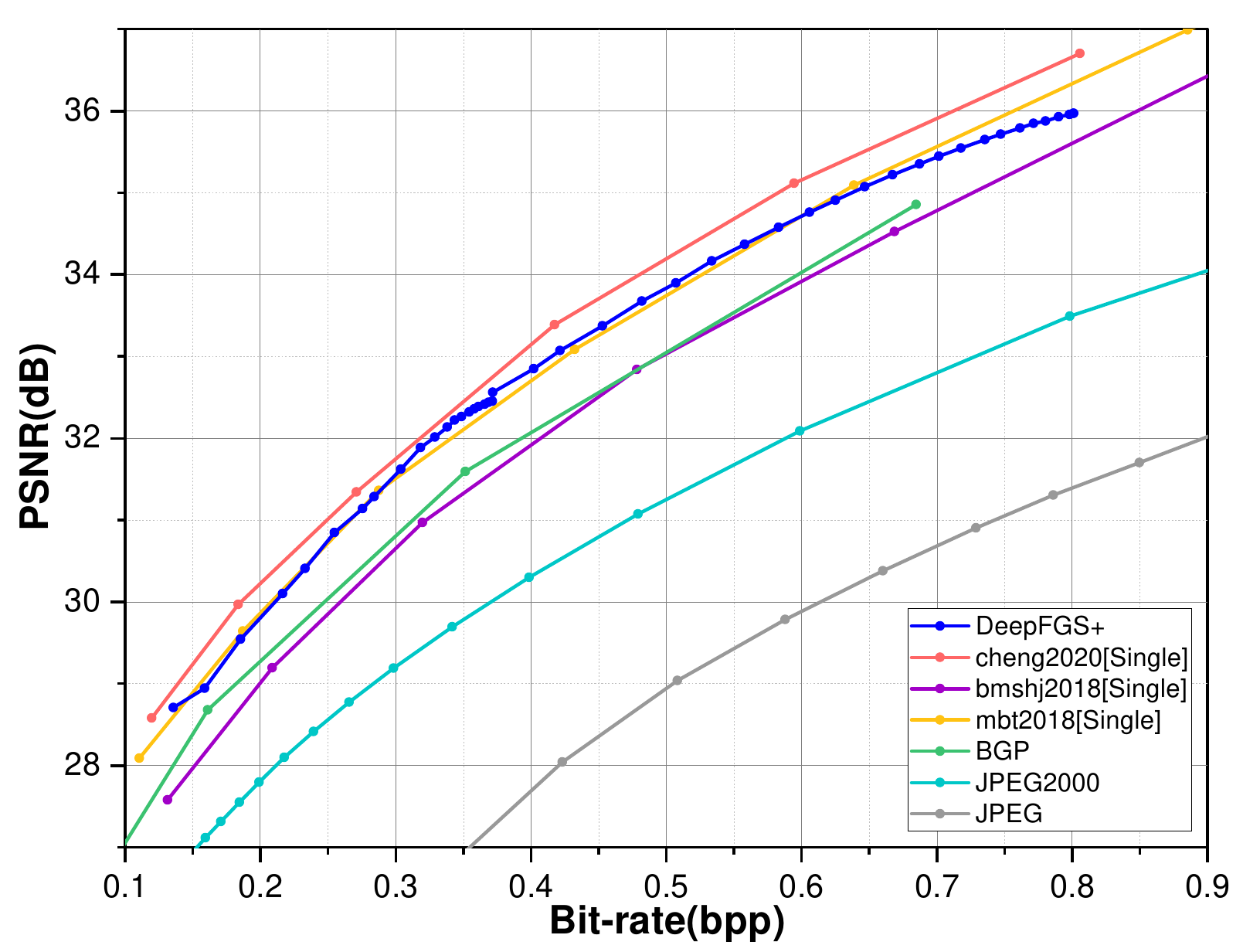}
        \caption{}
        \label{subfigure_3}
    \end{subfigure}
    \hspace{10mm}
  \begin{subfigure}[]{0.42\textwidth}
        \centering
      \includegraphics[width=1\textwidth]{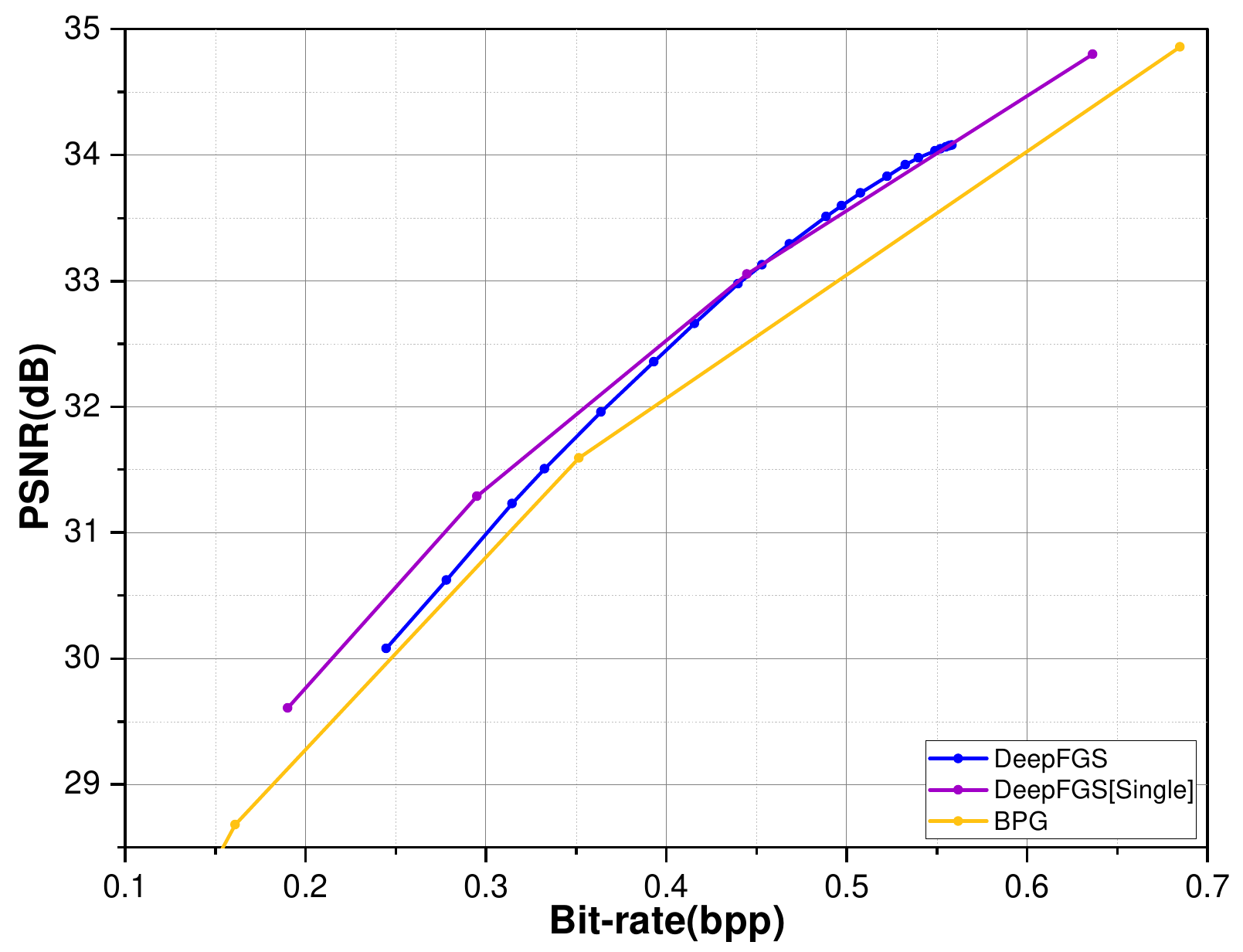}
        \caption{}
        \label{subfigure_4}
    \end{subfigure}
    
\caption{Comparison of rate-distortion performance of our DeepFGS with existing compression standards and other learned compression models, including Toderici \cite{toderici2015variable}, Zhang \cite{zhang2019learned}, Guo \cite{guo2019deep}, Jia \cite{jia2019layered}, Mei \cite{mei2021learning}, Cheng2020 \cite{cheng2020image}, bmshj2018 \cite{balle2018variational}, mbt2018 \cite{minnen2018joint}. (a): PSNR performance evaluation on Kodak dataset: compare with scalable coding models. (b): MS-SSIM performance evaluation on Kodak dataset: compare with scalable coding models. (c): PSNR performance evaluation on Kodak dataset: compare with non-scalable models. (d): PSNR performance evaluation on Kodak dataset: compare with non-scalable (single rate) DeepFGS.}
\label{fig:RMSE-100run}  
\vspace{-0.3cm}
\end{figure*}

\SubSection{Rate-distortion Performance}
\noindent\textbf{Compare with Scalable Coding Models.} We evaluate the rate-distortion performance on the Kodak image set \cite{kodak} with 24 uncompressed $768\times512$ images (more results on the high-definition data set will be given in the Appendix). The rate is measured by bits per pixel (bpp), and the quality is measured by PSNR and MS-SSIM. First, we compare performance with other scalable coding models. For traditional methods, SVC and HSVC, coarse-grained scalable coding mode performs better than fine-grained coding mode, so we compare with the former. As shown in Figure \ref{subfigure_1} and \ref{subfigure_2}, whether evaluated by PSNR or MS-SSIM, the rate-distortion performance of our model surpasses the existing traditional methods and learned methods by a big step. Furthermore, our model is also much more scalable than existing models. The bitstream of DeepFGS is nearly continuously adjustable so that DeepFGS can well decode the bitstream which is truncated at any position. For the convenience of comparison, in this section, we decode $y_s$ with 8 channels as the interval to draw the RD curve of our model. For the enlarged part in Figure \ref{subfigure_1}, this interval is 1 channel.

\noindent\textbf{Compare with Non-Scalable Models.} 
Due to the existence of redundancy between layers, compared to the non-scalable (single rate) model, the performance degradation brought by the traditional scalable method (SVC, SHVC) cannot be ignored. 
In order to explore the performance loss brought by scalability under the same network structure, we build a non-scalable model with a DeepFGS structure, called DeepFGS [Single]. 
As shown in Figure \ref{subfigure_4}, our DeepFGS model can overcome this disadvantage. The performance gap between the scalable model and the non-scalable model is very small. As shown in Figure \ref{subfigure_3}, the performance of DeepFGS is close to the cheng2020\cite{cheng2020image} of the non-scalable model.

\begin{figure}
  \centering
  \includegraphics[width=0.45\linewidth, trim = {0 20 0 10}, clip]{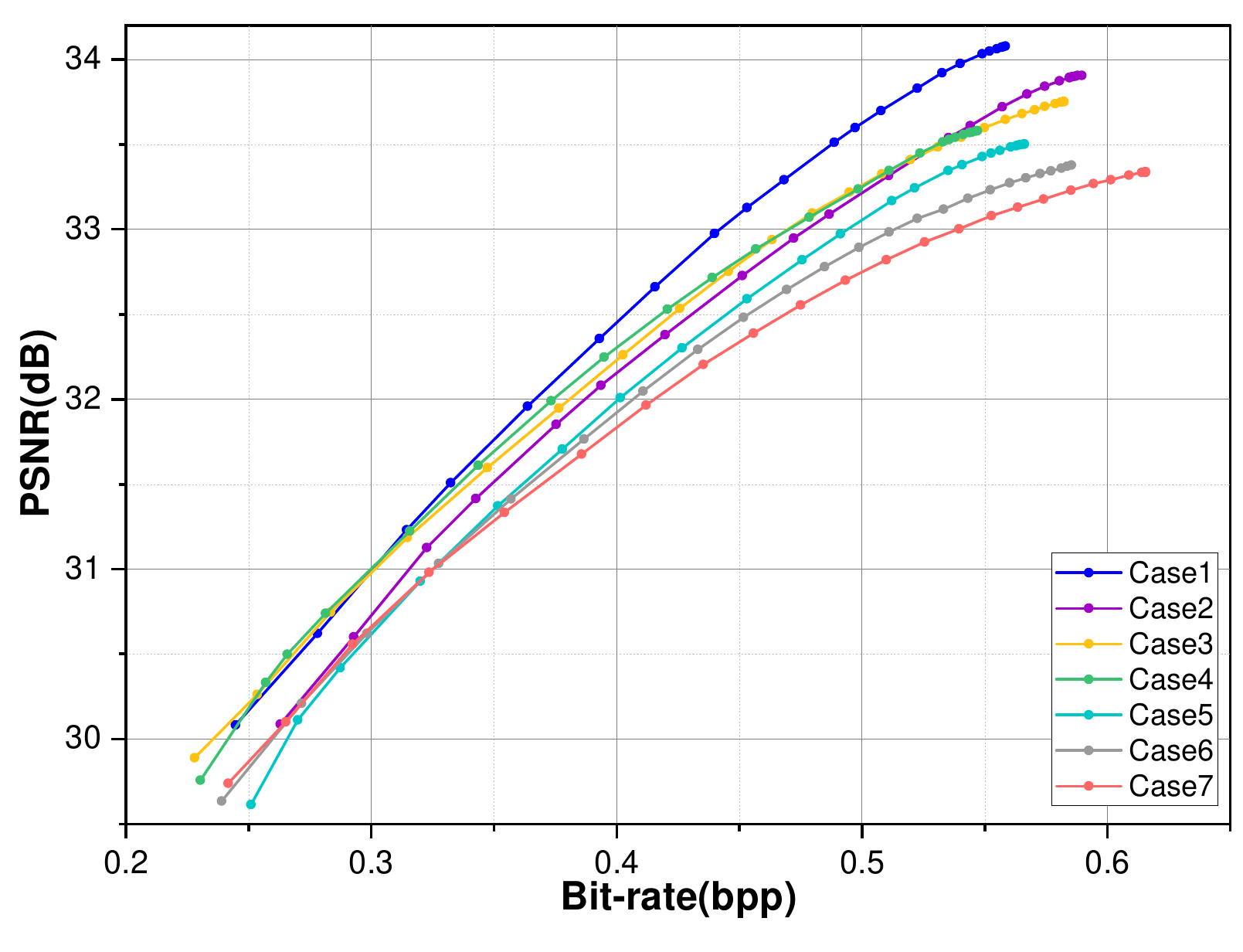}
  \caption{Ablation study on the proposed DeepFGS.}
  \label{ablation_pic}
\vspace{-0.2cm}
\end{figure}

\begin{table}
\begin{center}
\caption{The specific configuration of ablation study.}
\label{table1}
\resizebox{0.68\linewidth}{!}{
\begin{tabular}{cccccccc}
\toprule
 & {Case 1} & {Case 2} & {Case 3} & {Case 4} & {Case 5} & {Case 6} & {Case 7} \\ 
\midrule
{FRR} & \checkmark & \checkmark& \checkmark& \checkmark & & &\\
{FFM} & \checkmark & \checkmark& & & \checkmark  & &        \\
{MEM} & \checkmark &   &\checkmark & & & \checkmark &      \\

\bottomrule
\end{tabular}
}
\end{center}
\vspace{-1.5em}
\end{table}

\SubSection{Ablation Study}
\label{ablation}
To evaluate the contribution of each module to our model, we perform ablation studies shown in Figure \ref{ablation_pic}. We study the following models: feature-level redundancy reduction module (FRR), feature fusion module (FFM), and mutual entropy model (MEM). Table \ref{table1} shows the specific ablation configuration. The experimental results show that these modules all contribute to the performance improvement of DeepFGS, and combining them will achieve the best performance. In addition, by comparing the combinations (Case 2, Case 3) and (Case 4, Case 5), it can be found that MEM is more effective at high bitrates, while FFM is more effective at low bitrates.

\begin{figure*}
\begin{center}
\includegraphics[width=\linewidth]{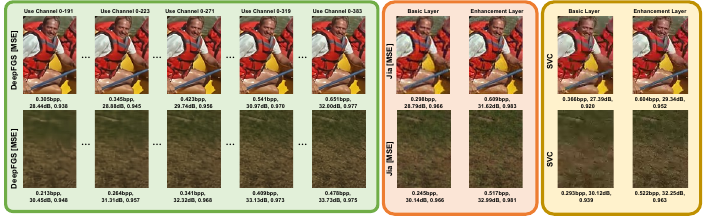}
\end{center}
\vspace{-0.3cm}
  \caption{Visual quality comparisons with the learning-based scalable method Jia \cite{jia2019layered} and traditional scalable codec SVC.}
\label{visual_results}
\vspace{-0.4cm}
\end{figure*}

\SubSection{Visual Results}
As shown in Figure \ref{visual_results}, we present some cropped reconstructed images on the Kodak dataset. Compared with traditional \cite{wiegand2007text} and learning-based \cite{jia2019layered} scalable coding models, our model has two main advantages. Firstly, our method is able to use fewer bitrates to generate more visually pleasing results. For example, in contour and textural regions (e.g. human face), our model offers much more structural details than other codecs. Secondly, our method can generate reconstructed images of different qualities in a fine-grained manner. As shown in the figure, our DeepFGS can provide dozens of continuous quality levels in the interval where \cite{wiegand2007text} and \cite{jia2019layered} can only generate two decoded images. 



\Section{Conclusion}
This paper proposes a learned fine-grained scalable coding model, namely DeepFGS, that achieves optimal compression performance and scalability. Ablation studies demonstrate the effectiveness of our proposed modules. In the following research, we will further explore the scalability of the image compression model, such as reducing the minimum tunable interval of the bitstream or increasing the rate range covered by the model.

\Section{Acknowledgment}
\sloppy{}
This work is financially supported for Outstanding Talents Training Fund in Shenzhen, this work is also supported by Shenzhen Science and Technology Program-Shenzhen Cultivation of Excellent Scientific and Technological Innovation Talents project (Grant No. RCJC20200714114435057), National Natural Science Foundation of China U21B2012.

\Section{References}
\bibliographystyle{IEEEbib}
\bibliography{refs}

\end{document}